# Domain Walls Conductivity in Hybrid Organometallic Perovskites: The Key of $CH_3NH_3PbI_3$ Solar Cell High Performance

Sergey N. Rashkeev[1*], Fedwa El-Mellouhi[1], Sabre Kais[1,2], and Fahhad H. Alharbi[1]

[1]Qatar Foundation, Qatar Environment and Energy Research Institute, P. O. Box 5825, Doha, Qatar

[2]Department of Chemistry, Birck Nanotechnology Center, Purdue University, West Lafayette, IN 47907, USA

*email: srashkeev@qf.org.qa

**The past several years has witnessed a surge of interest in organometallic trihalide perovskites, which are at the heart of the new generation of solid-state solar cells. Here, we calculated the static conductivity of charged domain walls in *n*- and *p*- doped organometallic uniaxial ferroelectric semiconductor perovskite $CH_3NH_3PbI_3$ using the Landau-Ginzburg-Devonshire (LGD) theory. We find that due to the charge carrier accumulation, the static conductivity may drastically increase at the domain wall by 3 – 4 orders of magnitude in comparison with conductivity through the bulk of the material. Also, a two-dimensional degenerated gas of highly mobile charge carriers could be formed at the wall. The high values of conductivity at domain walls and interfaces explain high efficiency in organometallic solution-processed perovskite films which contains lots of different point and extended defects. These results could suggest new routes to enhance the performance of this promising class of novel photovoltaic materials.**

Converting solar energy into electricity makes use of solar cells and is based on the photovoltaic effect. Single crystalline semiconductors such have been reported to achieve high power conversion efficiencies (25.6% for Si and 28.8% for GaAs) [1]. However, the manufacturing costs of single crystal solar cells are still relatively high so development of new technologies with lower production costs is of great interest. The past several years has witnessed a surge of interest in organometallic trihalide perovskites, which are at the heart of the new generation of solid-state solar cells [2-4]. In particular, methylammonium (MA) lead triiodide ($CH_3NH_3PbI_3$ or $MAPbI_3$) has attracted the most attention in mesoscopic solar cells [5,6]. The power conversion efficiency of these devices has shot up dramatically (and now exceeds 19%) [7], which is the result of favorable properties combination of $MAPbI_3$, including large absorption coefficient and long carrier diffusion lengths [5].

The structural, optical, and electronic properties of $MAPbI_3$ perovskites have been studied in great detail [8-12]. Perovskite $β$-$MAPbI_3$ belongs to the I4/mcm space group, and therefore, it is expected to be ferroelectric [13-15]. The current − voltage response of its bulk samples showed hysteretic behavior suggesting bulk ferroelectricity [13]. It was suggested that polarized ferroelectric domains within the $\beta$-MAPbI$_3$ film may act as small internal $p-n$ junctions, aiding the separation of photoexcited electron and hole pairs which reduces recombination [16,17]. Recently, dense $\beta$-MAPbI$_3$ thin films were investigated using piezoforce microscopy (PFM) [14] and the presence of relatively large (~100 nm) ferroelectric domains in solution-processed perovskite thin films were observed.

A presence of ferroelectric domains may explain some phenomena including the hysteresis in the current−voltage response [13,18] and giant dielectric constant [11]. Also, ferroelectric domains should significantly change the transport of charge carriers and their recombination rate because domain walls may exhibit static interface conductivity which is orders of magnitude higher than conductivity in the bulk of the semiconductor [19,20]. $\beta$-MAPbI$_3$ thin films used in solar cells are often mixed together with different meso- scale scaffolding materials [21], and additional interfaces between ferroelectric domains and non-ferroelectric scaffolds should also influence the charge carriers transport.

Recently, it was shown that both electron and hole diffusion lengths are at least 100 nm in solution processed CH$_3$NH$_3$PbI$_3$ [8], i.e., the high photoconversion efficiencies of these systems may stem from the comparable optical absorption and charge-carrier diffusion lengths. Another interesting feature is that the chemical and physical properties of organometallic trihalide perovskite materials strongly depend on the preparation method, and it may behave as $p$- or $n$-type semiconductors [13]. The presence of pre-existing charge carriers can also affect transport properties of the film. Using the electron and hole diffusion coefficients for MAPbI$_3$ (0.036 and 0.022 cm$^2$/s) obtained from fitting the decay dynamics of photoluminescence spectra to the diffusion model [8], one could estimate the mobility of the charge carriers using the Einstein's relation. At room temperature, one gets $\mu_e$ ~ 1.5 cm$^2$/(V·s) for the electron mobility which is significantly lower than the value of $\mu_e$ ~ 66 cm$^2$/(V·s) obtained from combined resistivity and Hall effect experimental data [13]. It means that simple diffusion of charge carriers through the bulk cannot account for the transport in this perovskite material.

In this context, we have investigated charge accumulation at ferroelectric domain walls in $n$- and $p$-doped $\beta$-MAPbI$_3$ thin films using the set of parameters taken from different experiments and existing first-principles calculations. MAPbI$_3$ is the organometallic trihalide material which has been investigated for a long time starting 1980s. Although other similar materials with the formula (RNH$_3$)MX$_3$ (R is an organic group, R=H-, CH$_3$-, NH$_3$CH-, etc.; M is Pb or Sn; and X is a halogen I, Br, or Cl) could also be perspective for solar cells, existing experimental and theoretical data for them are incomplete and sometimes contradictory. In tetragonal $\beta$-MAPbI$_3$ one should expect two types of domain walls (90$^o$ and 180$^o$ walls)) [22]. PFM studies show that ferroelectric domains in the $\beta$-MAPbI$_3$ thin films are clearly indicated by the complete 180° phase-contrast [14], so one could consider 180$^o$ charged domain walls in uniaxial ferroelectric [19] as a model for domain walls. Charged domain walls create strong electric field which causes free charge accumulation across the wall and sharply increase the domain-wall conductivity. In particular, a presence of meso-scale scaffolding (such as $TiO_2$, $Al_2O_3$, etc.) used in perovskite solar cells [21], provides tons of defects at which these walls could be pinned and exist in a steady state.

Following Ref. [19], let us first consider a head-to-head and tail-to-tail inclined wall in a uniaxial ferroelectric semiconductor doped with *n*-type impurity (for *p*- type doping the results are similar). A sketch of the charged walls is shown in Fig. 1. We suggest that the domain wall is planar. For the uniaxial ferroelectrics, the electric field potential $\varphi(\xi)$ and the ferroelectric polarization component $P_z(\xi)$ (order parameter) should be determined from the coupled Poisson and LGD equations (Supplementary), with boundary conditions of the potential vanishing far from the domain wall and the absolute value of $P_z(\xi)$ reaching the value of $P_S$. Several equations for concentration of different charged species (electrons, holes, and charged donors and acceptor impurities), for generation of the electron-hole pairs, and carriers recombination were also added.

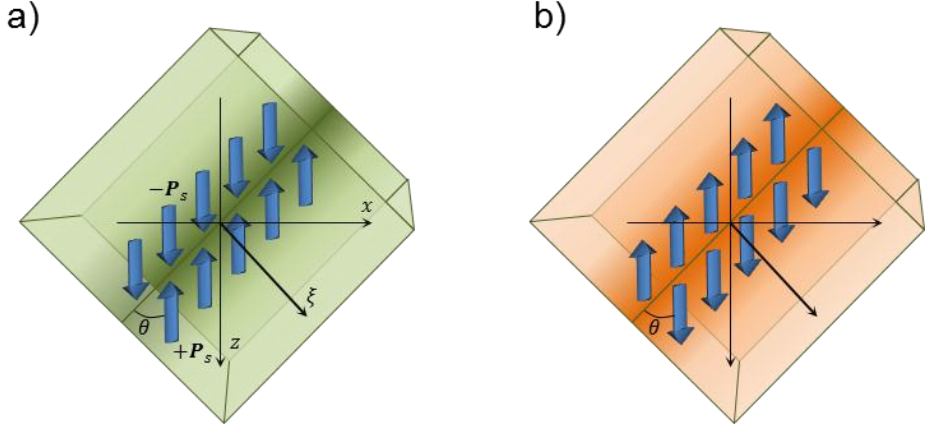

**Figure 1**. Sketch of the charged walls in the uniaxial ferroelectric semiconductors of *n* type: (a) inclined head-to-head, (b) inclined tail-to-tail domain walls. Green (orange) gradient color corresponds to excess negative (positive) charge density at the domain-wall vicinity. $\theta$ is the incline angle of the domain wall (the angle between the wall plane and the polarization vector of the uniaxial ferroelectric); the normal vectors to both film interfaces with electron and hole conductors are oriented along z axis; $\xi$ is the coordinate perpendicular to the wall; $P_S$ is the spontaneous ferroelectric polarization. For *n*- doped material, the excess negative charge is related to electrons while the excess positive charge – to both free-carrier holes and charged donor impurities.

Then the static conductivity can be calculated as (ions are less mobile species so their contribution to the conductivity is neglected),

$$\sigma(\xi) = q \cdot [\mu_e n(\xi) + \mu_p p(\xi)], \qquad (1)$$

where *q* is the electron charge, $n(\xi)$ and $p(\xi)$ are electrons and holes concentrations at a distance $\xi$ from the wall, $\mu_e$ and $\mu_p$ – the electron and hole mobility. To perform numerical calculations for domain walls, a significant number of different parameters is needed including dielectric function at different frequencies and polarizations, effective masses, coherence length, spontaneous polarization, etc. Fortunately, for the most investigated organometallic trihalide $MAPbI_3$, one has a significant amount of experimentally measured and calculated parameters which are sufficient for the LGD based modeling. A detailed and accurate description of all the used parameters is provided in Supplementary. In particular, we found that typical concentrations of charge carriers injected by solar light do not exceed $10^{13} - 10^{14}$ cm$^{-3}$ which is lower than typical concentrations of impurities in doped semiconductors.

The concentration of electrons $n(\xi)$ and charged donors $N_d^+(\xi)$ as a function of the distance from the wall plane $\xi$ (measured in the units of the correlation radius $r_c$), calculated for the inclined head-to-head domain walls with different slope angles $\theta$ and two different concentrations of donor impurities are shown in Figures 2(a) and 2(c). Detailed information about electric field and polarization distribution near the wall is provided in Supplementary. The net electric field of the bound charge attracts free electrons. The electron concentration is the highest for the perpendicular wall ($\theta = \pi/2$); it decreases with the bound charge decrease ($\theta$ decrease) and vanishes at $\theta = 0$. The net electric field "repulses" ionized donors (neutralizes them in the region with excess concentrations of electrons) and forms ionized donor depletion region.

The electric field in the vicinity of the wall is mainly defined by the concentration of degenerated electron gas which does not depend on the doping level. The wall accumulates electrons from nearby region in order to compensate the effects of the bound charges at the domain wall (and polarization field discontinuity). The electron attraction to this region naturally stops when the electron concentration reaches a saturation level which depends only on the value of spontaneous polarization field. For the head-to-head domain walls and $n$- doped material, the concentration of charged donors is depleted in the wall region because the probability for a charged donor defect to become neutral increases with the increase of concentration of free electrons. Calculations also indicated that the hole concentration in the vicinity of the wall is much lower than the concentration of electrons (see Supplementary, Fig. S4(a)), i.e., one can neglect hole related conductivity.

Figures 2 (c) and (d) also show the local-to-bulk conduction ratio, $\sigma/\sigma_{bulk}$. As a result of electron accumulation near the head-to-head domain wall, the static conductance drastically increases at the wall – up to 2 – 3 orders of magnitude for the dopant level of $N_{d0} = 10^{18}$ cm$^{-3}$ and up to 4 orders of magnitude for $N_{d0} = 10^{17}$ cm$^{-3}$. This is not surprising because the "saturated" value of the electron concentration at the wall does not depend on $N_{d0}$ while the electron concentration of the bulk is completely defined by $N_{d0}$ (i.e., saturated $n$ is smaller in the system with smaller $N_{d0}$). Also, conductance in the direction parallel to the wall is maximal for a perpendicular wall ($\theta = \pi/2$). For angle $\theta = 0$, there is no current across the film because this domain wall does not accumulate any charge. It is apparent that the ratio goes up for smaller $N_{d0}$ reaching the value of ~$10^4$ for $N_{d0} = 10^{17}$ cm$^{-3}$ and $\theta = \pi/2$.

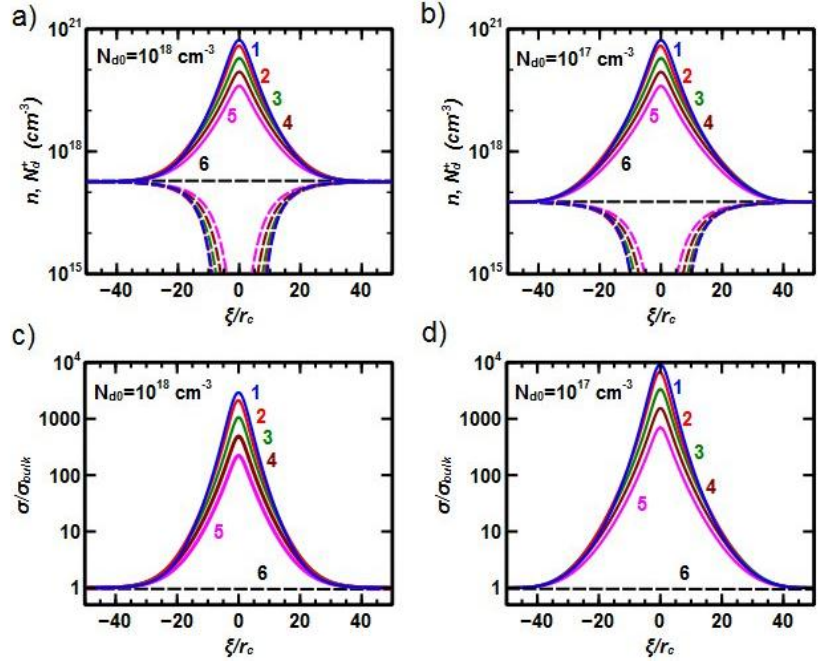

**Figure 2**. Concentrations of electrons $n(\xi)$ (solid lines) and ionized donors $N_d^+(\xi)$ (dashed lines), in the vicinity of the inclined head-to-head domain wall for concentration of $n$ donors in the bulk of material: (a) $N_{d0} = 10^{18}$ cm$^{-3}$; (b) $N_{d0} = 10^{17}$ cm$^{-3}$, and the values of local conductance to bulk conductance ratio $\sigma(\xi)/\sigma_{bulk}$ for: (c) $N_{d0} = 10^{18}$ cm$^{-3}$; (d) $N_{d0} = 10^{17}$ cm$^{-3}$, as functions of distance $\xi$ from the wall. All curves were calculated for different incline angles: $\theta = \pi/2; \pi/4; \pi/8; \pi/16; \pi/32; 0$ (curves 1 – 6); the distance $\xi$ is measured in the units of the coherence radius $r_c$.

For the tail-to-tail walls in $n$- doped semiconductors, the electric field potential at the tail-to-tail wall shows a very narrow and deep peak surrounded by a smooth, slowly growing "background" with the width strongly depending on $N_{d0}$ (Supplementary, Figs. S6(a) and S6(c)). The sharp dip corresponds to degenerated holes gas near the wall while the background is related to slow decay of the hole concentration to the saturated hole density at large distances which is several orders of magnitude lower than saturated electron density in $n$- doped semiconductor (Figs. 3(a) and 3(b)). Also, the background layer exhibits electron depletion due to electric field that pushes electrons away from both the dip and background regions.

Although there are differences in the behavior of positively (tail-to-tail) and negatively (head-to-head) charged domain walls, the highest charge carrier transport in both cases occurs in a thin layer near the wall. Also, each wall conducts only one type of charge carriers (electrons or holes), i.e., a significant spatial separation of charge carriers of different types takes place and should reduce the recombination of injected carriers. Holes accumulation near the tail-to-tail domain wall also drastically increases the static conductance at the wall (up to 4 orders of magnitude; Figs. 3(c) and 3(d)). At the tail-to-tail wall, however, the concentration of minority carriers (holes) significantly exceeds the concentration of majority carriers (electrons) only at the vicinity of the wall, i.e., holes conductance takes place only near the wall. The bulk (electron) conductance regime is reached at much larger distances from the wall. Therefore, tail-to-tail walls with high holes conductance are separated ("insulated") from the bulk material with electron conductance by a thicker electron depleted layer which should also reduce the possibility of recombination.

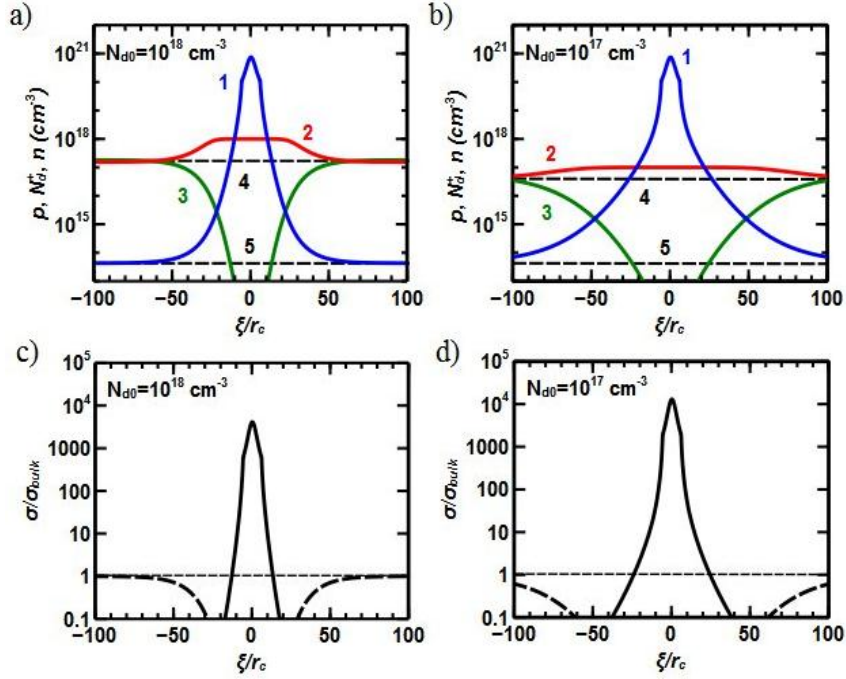

**Figure 3**. Dependencies of the hole density, $p(\xi)$ (blue curves 1); positively charged donor density $N_d^+$ (red curves 2), and; electron density, $n(\xi)$ (green curves 3) for the concentration of $n$ donors in the bulk: (a) $N_{d0} = 10^{18}$ cm$^{-3}$; (b) $N_{d0} = 10^{17}$ cm$^{-3}$. The incline angle is $\theta = \pi/4$. $\xi$ is measured in the units of the coherence radius $r_c$. Dashed lines 4 and 5 correspond to saturated values of the electron and hole densities at large distance from the tail-to-tail domain wall. The values of local conductance to bulk (electronic) conductance ratio for $\theta = \pi/4$, and: (c) $N_{d0} = 10^{18}$ cm$^{-3}$; (d) $N_{d0} = 10^{17}$ cm$^{-3}$ (the solid line relates to the holes conductance; dashed – to the corresponding electron contribution). $\xi$ is measured in the units of the coherence radius $r_c$.

Before now, we considered the case of $n$ doped semiconductor. The case of $p$ doped material should be considered in similar way by permuting electron and hole densities and taking into account the concentration of negatively charged acceptor impurities instead of $N_d^+$ considered above. It such case, head-to-head (tail-to-tail) walls are becoming holes (electrons) conductors.

Results presented in Figures 2(a) and 2(b) which show independence of the electron concentration at the domain wall on the doping level, were obtained in assumption that the size of the domain is much larger than the thickness of the domain wall, and charge accumulation at the wall does not cause significant electron depletion in the bulk of the domain where the electron concentration still remains close to the bulk concentration $n_\infty$. However, for a finite size domain, depletion should be taken into account. Assuming that the characteristic size of the domain is $L$ (the volume of the domain is $V \sim L^3$, and the surface area of the domain is $S \sim L^2$) and the thickness of the wall is $d \sim 15 \cdot r_c$, we find that the concentration of electrons at the wall cannot exceed the maximal value,

$$n_{max} \sim n_\infty \frac{V}{S \cdot d} \sim n_\infty \frac{L}{d} \qquad . \qquad (2)$$

It corresponds to the case when all electrons from the bulk of the domain moved to the domain walls. For MAPbI3 ($L \sim 100$ nm [14]; $r_c \sim 1.4$ nm [23]), $n_{max} \sim 5 n_\infty$, i.e., the concentration of accumulated electrons at the walls should be significantly lower than values of $10^3 - 10^4 n_\infty$ obtained in an assumption that domains are very large and depletion may be neglected.

For the minority charge carriers (holes) the situation is even worse. The holes concentration generated by natural solar light does not exceed $10^{13} - 10^{14}$ cm$^{-3}$. Therefore, in order to reach the quantum degenerated regime at the tail-to-tail wall with hole concentration of $10^{20} - 10^{21}$ cm$^{-3}$ (Figures 3(a) and 3(b)), one needs to collect holes from huge bulk volume which definitely exceeds any reasonable size of domains. Therefore, it seems unlikely to reach the degenerated quantum regime for holes.

These estimates provide possibility to evaluate an optimal domain size for different dopant densities. For heavily doped materials, the microstructure of the sample may be rougher (with a significant number of domain walls, interfaces, and grain boundaries) and still maintain a significant concentration of accumulated carriers at the domain walls to exhibit significant conductance. For undoped samples where most of the charge carriers are generated by solar light, and their concentrations are rather low, cleaner structures with larger domain sizes should be preferable.

Charge carrier accumulation at the domain walls is one of the factors that increase the conductance near the domain walls. However, in addition to carrier densities one should take into account the carrier mobility at the walls (Eq. (1)) which is related to many different factors. If the accumulated electron (hole) density at the domain wall reaches a certain value, a high-mobility degenerated two-dimensional electron gas (2DEG) (observed, e.g., in field effect transistors) may be created. Recently, it was shown that two-dimensional electron gas could be formed at the interface between perovskite SrTiO$_3$ and a spinel γ-Al$_2$O$_3$ epitaxial film with compatible oxygen ions sublattices [24]. The formed 2DEG exhibited electron Hall mobilities as large as $1.4 \times 10^5$ cm$^2$/Vs at 2K. Similar effects were observed at ferroelectric domain walls where the steady metallic-type conductivity reached $\sim 10^9$ times that of the parent matrix [25].

The electronic structure within the confinement was calculated by solving the Schroedinger equation for the plane-perpendicular direction in the potential shown in Supplementary (Figs.S2, S6). We found that 3 – 4 energy levels are contained in the dip of the potential energy before reaching the continuum. For qualitative consideration, one could start with considering only one plane-perpendicular electronic subband and neglect possible multiband effects. To evaluate the charge carrier mobility at the domain wall, one should consider several scattering mechanisms including: (i) phonon related intravalley scattering; (ii) phonon related intervalley g-process scattering; (iii) remote ionized impurity scattering, and; (iv) effects of possible interface (domain wall) roughness [26,27]. At low temperatures, the last two contributions should dominate. The concentration of positive donor impurities is low in the vicinity of the head-to-head wall (Figures 2(a) and 2(b)). Therefore, electrons flowing along the domain wall feel only the electrostatic field of remote positively charge donor impurities positioned near the edges of the conductive channel. The scattering, however, should be much stronger for holes at the tail-to-tail domain wall because the concentration of ionized impurities has a maximum near the wall (not minimum as for the head-to-head walls; Fig. 3).

Table 1 shows the parameters of two-dimensional charge carrier transport calculated in the Born approximation (see Supplementary) for different values of the incline angle $\theta$ and donor concentration $N_{d0} = 10^{17}$ cm$^{-3}$. The electron mobility at the head-to-head walls is extremely high and is comparable to the record electron mobility numbers obtained in a specially grown state of the art heterostructure (3.2×10$^7$ cm$^2$/Vs) at low temperatures [28]. It means that the 2DEG mobility at charged ferroelectric domain walls of the MAPbI$_3$ perovskite may be very high.

**Table 1**. Two-dimensional electron (hole) density in the accumulation layer, $N_s$; corresponding two-dimensional charge carrier (electrons or holes) gas Fermi wavelength, $k_F$; the transport relaxation time at the Fermi energy, $\tau(E_F)$; and electron (hole) mobility, $\mu$, for different values of the incline angle $\theta$ and donor concentration $N_{d0} = 10^{17}$ cm$^{-3}$.

| Incline angle $\theta$ | $\pi/2$ | $\pi/4$ | $\pi/8$ | $\pi/16$ | $\pi/32$ |
|---|---|---|---|---|---|
| $N_s$, 10$^{14}$ cm$^{-2}$ | 4.742$^{(a)}$ / 9.137$^{(b)}$ | 3.347 / 6.351 | 1.806 / 3.258 | 0.918 / 1.487 | 0.460 / 0.514 |
| $k_F$, 10$^7$ cm$^{-1}$ | 5.458 / 7.577 | 4.586 / 6.317 | 3.369 / 4.524 | 2.401 / 3.057 | 1.701 / 1.798 |
| $\tau(E_F)$, 10$^{-9}$ s | 2.960 / 0.189 | 1.768 / 0.114 | 0.694 / 0.045 | 0.233 / 0.015 | 0.074 / 0.003 |
| $\mu$, 10$^6$ cm$^2$/(Vs) | 32.527 / 1.038 | 19.428 / 0.628 | 7.621 / 0.249 | 2.557 / 0.083 | 0.815 / 0.019 |

$^{(a)}$Calculated for the head-to-head wall (charge carriers – electrons);
$^{(b)}$Calculated for the tail-to-tail wall (charge carriers – holes).

Table 1 clearly indicates that scattering by ionized impurities is much stronger for holes at the tail-to-tail domain walls than for electrons at the head-to-head walls in *n*- doped material. These numbers for the mobility were obtained in an assumption that the electron (hole) gas at the domain wall reaches a quantum state and becomes degenerated. However, it may happen that degeneracy won't be reached. The main factor which regulates the maximal accumulated electron density is the relation between the average size of ferroelectric domain and the thickness of the wall which separates a domain from a neighboring domain and/or from a grain of other material used for scaffolding (e.g., $TiO_2$).

In conclusion, we calculated the static conductance of charged head-to-head and tail-to-tail domain walls with different incline angles with respect to the spontaneous polarization vector in the *n*- and *p*-doped organometallic uniaxial ferroelectric semiconductor perovskite $CH_3NH_3PbI_3$, in which ferroelectric domains were observed directly using piezoforce microscopy. We showed that the carriers diffuse to domain walls and accumulate there. Due to this accumulation, the static conductance at the domain walls may increase by 3 – 4 orders of magnitude in comparison with conductance through the bulk of the material. These results support the transport mechanism suggested in Ref. [16] where it was suggested that the carriers diffuse along these 'ferroelectric highways' toward the electrodes, unimpeded by carriers of the opposite charge. Calculations of the charge carriers mobility confirm that this mechanism could be the reason for the origin of exceptionally long carrier diffusion lengths [8,29] despite local structural disorder. The high values of domain wall and interfacial conductance may be the main reason of the high efficiency in organometallic solution-processed perovskite solar cells which contain lots of different point and extended defects. These investigations could shed light on the fundamental photovoltaic mechanisms for perovskite-based solar cells and develop other materials based on domain walls conductance.

**References:**

[1] Green, M. A., Emery, K., Hishikawa, Y., Warta, W. & Dunlop, E. D. Solar cell efficiency tables (version 39). *Progress in Photovoltaics: Research and Applications* **20**, 12–20 (2012).

[2] Mcgehee, M. D. Materials science: Fast-track solar cells. *Nature* **501**, 323−325 (2013).

[3] Hodes, G. Perovskite-based solar cells. *Science* **342**, 317− 318 (2013).

[4] Service, R. F. Perovskite solar cells keep on surging. *Science* **344**, 458 (2014).

[5] Kim, H. S., Im, S. H. & Park, N.-G. Organolead halide perovskite: New horizons in solar cell research. *J. Phys. Chem. C* **118**, 5615−5625 (2014).

[6] Snaith, H. J. Perovskites: The emergence of a new era for low-cost, high-efficiency solar cells. *J. Phys. Chem. Lett*. **4**, 3623−363 (2013).

[7] Zhou, H. *et al*. Interface engineering of highly efficient perovskite solar cells. *Science* **345**, 542−546 (2014).

[8] Xing, G. *et al.* Long-range balanced electron and hole-transport lengths in organic-inorganic $CH_3NH_3PbI_3$. *Science* **342**, 344−347 (2013).

[9] Wolf, S. D. *et al*. Organometallic halide perovskites: Sharp optical absorption edge and its relation to photovoltaic performance. *J. Phys. Chem. Lett*. **5**, 1035−1039 (2014).

[10] Savenije, T. J. *et al*. Thermally activated exciton dissociation and recombination control the organometal halide perovskite carrier dynamics. *J. Phys. Chem. Lett*. **5**, 2189−2194 (2014).

[11] Juárez-Pérez, E. J. *et al*. Photoinduced giant dielectric constant in lead halide perovskite solar cells. *J. Phys. Chem. Lett*. **5**, 2390−2394 (2014).

[12] Tan, Z.-K. *et al*. Bright light-emitting diodes based on organometal halide perovskite. *Nature Nanotechnol*. **9**, 687−692 (2014).

[13] Stoumpos, C. C., Malliakas, C. D. & Kanatzidis, M. G. Semiconducting tin and lead iodide perovskites with organic cations: Phase transitions, high mobilities, and near-infrared photoluminescent properties. *Inorg. Chem*. **52**, 9019−9038 (2013).

[14] Kutes, Y. *et al*. Direct observation of ferroelectric domains in solution-processed $CH_3NH_3PbI_3$ perovskite thin films. *J. Phys. Chem. Lett*. **5**, 3335−3339 (2014).

[15] Lines, M. E. & Glass, A. M. *Principles and Applications of Ferroelectrics and Related Materials* (Oxford Univ. Press, London, 2001).

[16] Frost, J. M., Butler, K. T., Brivio, F., Hendon, C. H., Schilfgaarde, M. & Walsh, A. Atomistic origins of high-performance in hybrid halide perovskite solar cells. *Nano Lett*. **14**, 2584−2590 (2014).

[17] Frost, J. M., Butler, K. T. & Walsh, A. Molecular ferroelectric contributions to anomalous hysteresis in hybrid perovskite solar cells. *APL Mater*. **2**, 081506 (2014).

[18] Snaith, H. J. *et al*. Anomalous hysteresis in perovskite solar cells. *J. Phys. Chem. Lett*. **5**, 1511−1515 (2014).

[19] Eliseev, E. A., Morozovska, A. N., Svechnikov, G. S., Gopalan, V. & Shur, V. Ya. Static conductivity of charged domain walls in uniaxial ferroelectric semiconductors. *Phys. Rev. B* **83**, 235313 (2011).

[20] Vasudevan, R. K. *et al*. Domain wall conduction and polarization-mediated transport in ferroelectrics. *Adv. Funct. Mater.* **23**, 2592–2616 (2013).

[21] Edri, E. *et al*. Why lead methylammonium tri-iodide perovskite-based solar cells require a mesoporous electron transporting scaffold (but not necessarily a hole conductor). *Nano Lett*. **14**, 1000−1004 (2014).

[22] Damjanovic, D.. Ferroelectric, dielectric and piezoelectric properties of ferroelectric thin films and ceramics. *Rep. Prog. Phys*. **61**, 1267–1324 (1998).

[23] Choi, J. J. *et al*. Structure of methylammonium lead iodide within mesoporous titanium dioxide: active material in high-performance perovskite solar cells. *Nano Lett*. **14**, 127–133 (2014).

[24] Chen, Y. Z. *et al*. A high-mobility two-dimensional electron gas at the spinel/perovskite interface of $\gamma$-$Al_2O_3$/$SrTiO_3$. *Nature Comm*. **4**, 1371-1376 (2013).

[25] Sluka, T., Tagantsev, A. K., Bednyakov, P. & Setter, N. Free-electron gas at charged domain walls in insulating $BaTiO_3$. *Nature Comm*. **4**, 1808-1813 (2013).

[26] Ando, T., Fowler, A. B. & Stern, F. Electronic properties of two-dimensional systems. *Rev. Mod. Phys*. **54**, 437-672 (1982).

[27] Tanaka, T. *et al*. Experimental and theoretical analysis of the temperature dependence of the two-dimensional electron mobility in a strained Si quantum well. *J. Appl. Phys*. **111**, 073715 (2012).

[28] Kumar, A. et al. Nonconventional Odd-Denominator Fractional Quantum Hall States in the Second Landau Level. *Phys. Rev. Lett*. **105**, 246808 (2010).

[29] Stranks, S. D. *et al*. Electron-hole diffusion lengths exceeding 1 micrometer in an organometal trihalide perovskite absorber. *Science* **342**, 341–344 (2013).

**Author contributions**
S.N.R and F.H.A. conceived the experiment. S.N.R., F.E-M., and S.K. designed the simulation strategy. S.N.R. and F.E-M. performed theoretical calculations. S.N.R., F.H.A., and S.K. wrote the paper. All authors contributed to the scientific planning and discussions.

**Additional information**
Supplementary information is available in the online version of the paper. Correspondence and requests for materials should be addressed to S.N.R. or F.H.A.

**Competing financial interests**
The authors declare no competing financial interests.